\documentclass{amsart}
\usepackage{graphicx}

\theoremstyle{definition}

\theoremstyle{remark}

\numberwithin{equation}{section}

\begin{document}

\title{Improved Approximations of Hedges' g*}
\author{Xiaohuan Xue}
\address{Department of Mathematics and Statistics, Wake Forest University, Winston-Salem, North Carolina 27109}
\curraddr{Department of Psychology, University of California, Los Angeles, California 90095}
\email{xhxue@g.ucla.edu}

%    General info


\begin{abstract}
Hedges' unbiased estimator g* has been broadly used in statistics. We propose a sequence of polynomials to better approximate the multiplicative correction factor of g* by incorporating analytic estimations to the ratio of gamma functions.  
\end{abstract}

\maketitle
\section{Introduction}

Hedges proposed the widely used unbiased estimator, $g*$, of standardized mean differences \cite{Heges, Hedges Text, Intro}. Suppose group $i$ with sample size $n_i$ is $y_{i1},\cdots, y_{in_i}$, and group $j$ with sample size $n_j$ are $y_{j1},\cdots, y_{in_j}$, and we assume the samples in two groups are normally distributed with the same variance, that is,
 \begin{equation}
  y_{i1},\cdots, y_{in_i}\sim N(\mu_i,\sigma^2)
 \end{equation}
  \begin{equation}
  y_{j1},\cdots, y_{jn_j}\sim N(\mu_j,\sigma^2)
 \end{equation}

Hedges' $g*$ is given by the sample mean difference divided by a multiple of the pooled sample standard deviation as following

\begin{equation}
g*=J(n_i+n_j-2)\cdot\frac{\overline{y}_j-\overline{y}_i}{s_{p}}
\end{equation}
where
\begin{equation}
 J(m)=\frac{\Gamma(\frac{m}{2})}{\sqrt{\frac{m}{2}}\Gamma(\frac{m-1}{2})}
\end{equation}
and $s_{p}$ is the pooled sample standard deviation 
\begin{equation}
s_{p}=\sqrt{\frac{(n_j-1)s_j^2+(n_i-1)s_i^2}{n_j+n_i-2}}
\end{equation}

The multiplicative correction term $J(\cdot)$ in Hedges' $g*$ is not easy to calculate in practice, and one commonly used approximation is given by Hedges \cite{Heges, Hedges Text}
\begin{equation}
J(m)=\frac{\Gamma(\frac{m}{2})}{\sqrt{\frac{m}{2}}\Gamma(\frac{m-1}{2})}\approx 1-\frac{3}{4m-1}
\end{equation}
and let us denote Hedges' estimation as $H(m)$:
\begin{equation}
H(m)= 1-\frac{3}{4m-1}
\end{equation}
We now propose a sequence of polynomials to give more accurate approximations of $J(\cdot)$ and thus improve the accuracy of Hedges' $g*$.

\section{Wallis Ratio and Approximation of Hedge's Estimator}

The Wallis ratio \cite{Chu, M, Dum} is defined by

\begin{equation}
\frac{\Gamma(x+1)}{\Gamma(x+\frac{1}{2})}
\end{equation}
and we can see
\begin{equation}
    \sqrt{\frac{m}{2}}J(m)=\frac{\Gamma(\frac{m}{2})}{\Gamma(\frac{m-1}{2})}
\end{equation} 
is a special case of Wallis ratio when $x=\frac{m}{2}-1$. Thus the approximations of Wallis ration as a result of properties of gamma function can be applied to estimating $J(\cdot)$.

For Wallis ratio, there are Mortici's approximations \cite{M, Dum, Paris}

\begin{equation}
 \frac{\Gamma(x+1)}{\Gamma(x+\frac{1}{2})}\approx \sqrt{x+\frac{1}{4}}
\end{equation}

\begin{equation}
 \frac{\Gamma(x+1)}{\Gamma(x+\frac{1}{2})}\approx \sqrt[4]{x^2+\frac{1}{2}x+\frac{1}{8}}
\end{equation}

\begin{equation}
 \frac{\Gamma(x+1)}{\Gamma(x+\frac{1}{2})}\approx \sqrt[6]{x^3+\frac{3}{4}x^2+\frac{9}{32}x+\frac{5}{128}}
\end{equation}

\begin{equation}
 \frac{\Gamma(x+1)}{\Gamma(x+\frac{1}{2})}\approx \sqrt[8]{x^4+x^3+\frac{1}{2}x^2+\frac{1}{8}x}
\end{equation}

\begin{equation}
 \frac{\Gamma(x+1)}{\Gamma(x+\frac{1}{2})}\approx \sqrt[10]{x^5+\frac{5}{4}x^4+\frac{25}{32}x^3+\frac{35}{128}x^2+\frac{75}{2048}x+\frac{3}{8192}}
\end{equation}

\begin{equation}
 \frac{\Gamma(x+1)}{\Gamma(x+\frac{1}{2})}\approx \sqrt[12]{x^6+\frac{3}{2}x^5+\frac{9}{8}x^4+\frac{1}{2}x^3+\frac{15}{128}x^2+\frac{3}{256}x+\frac{11}{1024}}
\end{equation}

By letting $x=\frac{m}{2}-1$, we can have an approximation of $J(m)$
\begin{equation}
 J(m)\approx P_1(m)=\sqrt{1-\frac{3}{2m}}
\end{equation}

\begin{equation}
\begin{split}
 J(m)\approx P_2(m)&=\sqrt{\frac{2}{m}}\sqrt[4]{(\frac{m}{2}-1)^2+\frac{1}{2}(\frac{m}{2}-1)+\frac{1}{8}}\\&=\sqrt[4]{1-\frac{3}{m}+\frac{5}{2m^2}} 
\end{split}
 \end{equation}

\begin{equation}
\begin{split}
  J(m)\approx P_3(m)&=\sqrt{\frac{2}{m}} \sqrt[6]{(\frac{m}{2}-1)^3+\frac{3}{4}(\frac{m}{2}-1)^2+\frac{9}{32}(\frac{m}{2}-1)+\frac{5}{128}} \\
  &=\sqrt[6]{1-\frac{9}{2m}+\frac{57}{8m^2}-\frac{63}{16m^3}}
\end{split}
\end{equation}

\begin{equation}
\begin{split}
   J(m)\approx P_4(m)&=\sqrt{\frac{2}{m}}\sqrt[8]{(\frac{m}{2}-1)^4+(\frac{m}{2}-1)^3+\frac{1}{2}(\frac{m}{2}-1)^2+\frac{1}{8}(\frac{m}{2}-1)}  \\
   &= \sqrt[8]{1-\frac{6}{m}+\frac{14}{m^2}-\frac{15}{m^3}+\frac{6}{m^4}}
\end{split}
\end{equation}

\begin{equation}
\begin{split}
 J(m)&\approx P_5(m)\\
 &=\sqrt{\frac{2}{m}} \sqrt[10]{(\frac{m}{2}-1)^5+\frac{5}{4}(\frac{m}{2}-1)^4+\frac{25}{32}(\frac{m}{2}-1)^3+\frac{35}{128}(\frac{m}{2}-1)^2+\frac{75}{2048}(\frac{m}{2}-1)+\frac{3}{8192}}    \\
 &=\sqrt[10]{1-\frac{15}{2m}+\frac{185}{8m^2}-\frac{585}{16m^3}+\frac{3755}{128m^4}-\frac{2409}{256m^5}}
\end{split}
\end{equation}
\begin{equation}
 J(m)\approx P_6(m)=\sqrt[12]{1-\frac{9}{m}+\frac{69}{2m^2}-\frac{72}{m^3}+\frac{687}{8m^4}-\frac{441}{8m^5}+\frac{247}{16m^6}}
\end{equation}

\section{Accuracy of different approximations}

In this section, we compare the accuracy of all the given approximations by measuring the absolute values of their errors to the real value with a broad range of $m$. For convenience, we introduce the notations for the absolute errors

\begin{equation}
    \delta_0 (m)=|H(m)-J(m)|
\end{equation}
\begin{equation}
    \delta_i (m)=|P_i(m)-J(m)|; i=1,\cdots,6.
\end{equation}

\begin{figure}[tb]
\includegraphics[scale=0.5]{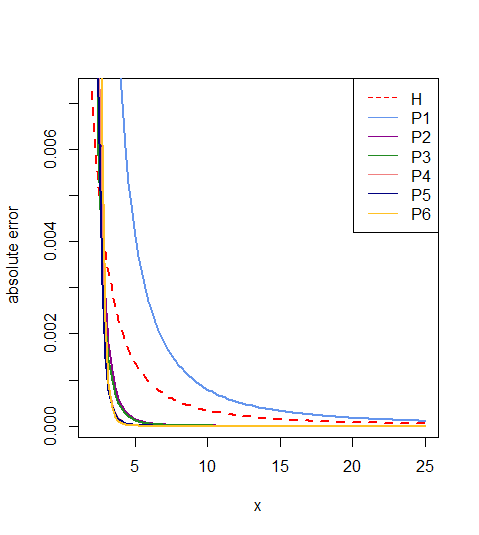}
\caption{The approximation accuracy of $H$ and $P_n$.}
\label{fig1}
\end{figure}

Figure ~\ref{fig1} shows the performance of approximations and we can see Hedges' approximation $H(m)$, which is the dashed red line, has less accuracy compared to $P_2(m),P_3(m),\cdots,P_m(m)$. In terms of absolute errors, we can see 
\begin{equation}
    \delta_6 (m)< \delta_5 (m)< \delta_4 (m)< \delta_3 (m)< \delta_2 (m)< \delta_0 (m)< \delta_1 (m).
\end{equation}

This can also be verified by performing numerical analysis demonstrated by Table ~\ref{FIGtable}, from which the order of accuracy is more straightforward.  

\begin{table}[ht]
\caption{Numerical accuracy of different approximations}\label{FIGtable}
\noindent\[
\begin{array}{c c c c c c c c}
\hline
m & \delta_0 & \delta_1 & \delta_2 & \delta_3 & \delta_4 & \delta_5& \delta_6\\
\hline
10 & 0.00033 & 0.00079 & 5.34e-06 & 4.66e-06 & 1.43e-07 & 1.35e-07 & 8.47e-09
\\
30&  3.55e-05& 7.49e-05 & 4.60e-08& 4.03e-08& 1.13e-10&1.067e-10 &6.24e-13\\
50 & 1.27e-05 & 2.62e-05 &5.56e-09 &4.86e-09 & 4.71e-12 & 4.45e-12 & 1.79e-14\\
70& 6.44e-06& 1.32e-05 &1.40e-09 & 1.23e-09 &5.93e-13 &5.60e-13 &5.66e-15\\
100 & 3.15e-06 & 6.39e-06 & 3.29e-10 &2.88e-10 & 4.49e-14 & 4.11e-14 &  2.31e-14 \\
200&7.84e-07 &1.58e-06 &2.00e-11 &1.75e-11 &4.29e-14 &4.29e-14 & 4.19e-14\\
\hline
\end{array}
\]
\end{table}

\section{discussion}

In this paper, we have proposed a sequence of more accurate approximations to the multiplicative correction factor, $J(m)$, in Hedges' unbiased estimator of standardized mean difference.

\begin{figure}[tb]
\includegraphics[scale=0.5]{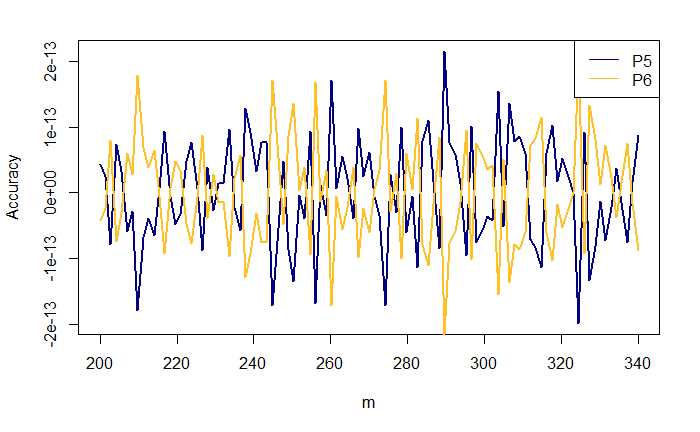}
\caption{The approximation accuracy of $P_5$ and $P_6$ when $m>200$.}
\label{fig2}
\end{figure}

It is also worth mentioning that the difference between $P_5(m)$ and $P_6(m)$ are small when $m$ is over 100, and there is almost no difference when $m$ is over 200. We can also see this from Figure ~\ref{fig2}, both  $P_5(m)$ and $P_6(m)$ are osculating around 0 within the magnitude of $2\times 10^{-13}$.

More accurate and efficient approximations to Hedges' $g*$ would be available with deeper understandings of the properties of gamma functions, which has always been an appealing topic in both mathematics itself and our future work.

\bibliographystyle{amsplain}

\end{document}